\documentclass[journal,comsoc]{IEEEtran}

\usepackage{graphicx}
\usepackage{epstopdf}
\usepackage{float}
\usepackage{algorithm,algorithmic}
\usepackage{array}
\usepackage{amsmath}
\usepackage{amssymb}
\usepackage{mdwmath}
\usepackage{mdwtab}
\usepackage{eqparbox}

\usepackage{fixltx2e}
\usepackage{cases}
\usepackage{bm}
\usepackage{multirow}
\usepackage{psfrag}
\usepackage[usenames]{color}
\usepackage{mathtools}

\usepackage{fixmath}
\usepackage{mathdots}
\usepackage{latexsym}
\usepackage{amsmath,amssymb}

\usepackage{bm}
\usepackage{footmisc}

\usepackage[T1]{fontenc}
\usepackage{url}
\usepackage{cite}
\newcommand{\subparagraph}{}
\usepackage{titlesec}

\newcommand{\diag}{\mathop{\mathrm{diag}}}
\newcommand{\tr}{\mathop{\mathrm{Tr}}}

\begin{document}

\title{Signal Shaping for Generalized Spatial Modulation and Generalized Quadrature Spatial Modulation}
\author{Shuaishuai~Guo,~\IEEEmembership{Member, IEEE,}
Haixia~Zhang,~\IEEEmembership{Senior Member, IEEE,}
Peng Zhang,~\IEEEmembership{Member, IEEE,}\\
Shuping Dang,~\IEEEmembership{Member, IEEE,} Cong Liang, and   Mohamed-Slim Alouini,~\IEEEmembership{Fellow, IEEE}
\thanks{S. Guo, S. Dang and M. -S. Alouini are with Department of Electrical Engineering, Computer
Electrical and Mathematical Sciences \& Engineering (CEMSE) Division, King
Abdullah University of Science and Technology (KAUST), Thuwal, Makkah
Province, Kingdom of Saudi Arabia, 23955-6900 (email: \{shuaishuai.guo; shuping.dang; slim.alouini\}@kaust.edu.sa).}
\thanks{H. Zhang and C. Liang are with Shandong Provincial Key Laboratory of Wireless Communication Technologies, Shandong University, Jinan, China, 250061 (e-mail: haixia.zhang@sdu.edu.cn; congliang@mail.sdu.edu.cn).}
\thanks{P. Zhang is with the School of Computer Engineering, Weifang University, Weifang 261061, China (e-mail: sduzhangp@163.com).}
\thanks{Manuscript received on Dec. XX, 2018}}
\markboth{Submitted to IEEE TRANSACTIONS ON WIRELESS COMMUNICATIONS}%
{Guo \MakeLowercase{\textit{et al.}}:Signal Shaping for Generalized Spatial Modulation and Generalized Quadrature Spatial Modulation}
\maketitle
\begin{abstract}
This paper investigates generic signal shaping methods for multiple-data-stream generalized spatial modulation (GenSM) and generalized quadrature spatial modulation (GenQSM) based on the maximizing the minimum Euclidean distance (MMED) criterion. Three cases with different channel state information at the transmitter (CSIT) are considered, including no CSIT, statistical CSIT and perfect CSIT. A unified optimization problem is formulated to find the optimal transmit vector set under size,  power  and   sparsity constraints.  We propose an optimization-based signal shaping (OBSS) approach by solving the formulated problem directly  and a codebook-based signal shaping (CBSS) approach by finding  sub-optimal solutions in discrete space. In the OBSS approach, we reformulate the original problem to optimize the signal constellations used for each transmit antenna combination (TAC).  Both the size and entry of all signal constellations are optimized. Specifically, we suggest the use of a recursive design for size optimization. The entry optimization is formulated as a non-convex large-scale quadratically constrained quadratic programming (QCQP) problem and can be solved by existing optimization techniques with rather high complexity. To reduce the complexity, we propose the CBSS approach using a codebook  generated by quadrature amplitude modulation (QAM) symbols and a low-complexity selection algorithm to choose the optimal transmit vector set. Simulation results show that the OBSS approach exhibits the optimal performance in comparison with existing benchmarks. However, the OBSS approach is impractical for large-size signal shaping and adaptive signal shaping with instantaneous CSIT due to the demand of high computational complexity. As a low-complexity approach, CBSS shows comparable performance and can be easily implemented in large-size systems.
\end{abstract}

\begin{IEEEkeywords}
Multiple-input multiple-output, generalized spatial modulation, generalized quadrature spatial modulation,  signal shaping, constellation design, precoding, maximize the minimum Euclidean distance, sparsity constraint
\end{IEEEkeywords}

\IEEEpeerreviewmaketitle

\section{Introduction}
\IEEEPARstart{M}{ultiple-data-stream} generalized spatial modulation (GenSM)  and generalized quadrature spatial modulation (GenQSM)  have emerged as new techniques for multiple-input multiple-output (MIMO) communications with reduced radio frequency (RF) chains and fast antenna switches. As generalized forms of spatial modulation (SM) \cite{Mesleh2008,Yang2008,Renzo2014,Yang2015} and quadrature spatial modulation (QSM) \cite{Mesleh2015,Li2017,Mohaisen2018}, they employ multiple in-phase and quadrature (IQ) RF chains for multiple-data-steam transmission and additionally carry information by the selection of transmit antenna combinations (TACs). 
By varying the number of RF chains, they can achieve an attractive trade-off between spectral efficiency (SE) and energy efficiency (SE).
For  general $(N_t,N_r,N_{RF},n)$ GenSM/GenQSM MIMO systems with $N_t$ transmit antennas and $N_r$ receive antennas conveying a fixed length of $n$-bit stream via $N_{RF}$ RF chains,  this paper investigates generic signal shaping methods to find the optimal $2^n$ transmit vectors. This is a rather intricate task because it couples the multiple-dimensional signal constellation optimization as well as  the spatial constellation optimization. In this paper, we aim to solve the problem based on the maximizing the minimum Euclidean distance (MMED) criterion. According to \cite{Payaro2009,Vu2007,Wang2016}, the MMED criterion is equivalent to the criteria of minimizing the symbol error rate (MSER) and maximizing the mutual information (MMI) in the high signal-to-noise ratio (SNR) regime.
\subsection{Prior Work}
All existing signal constellation designs, spatial constellation designs, precoding schemes (including phase rotation and power allocation schemes) as well as their combinations  can be viewed as signal shaping methods, because all of them affect the transmit vectors.
Based on the required information, we classify prior work into the following two categories:

\subsubsection{Signal Shaping without CSIT}
Without CSIT, GenSM/GenQSM systems can benefit from the off-line design of signal and spatial constellation. Specifically, \cite{Yang2014,Zhang2015,Maleki2015} investigated the signal constellation design for SM based on pre-defined constellation structures or optimization techniques. However,  the assumption that all TACs utilize the same signal constellation in \cite{Yang2014,Zhang2015,Maleki2015} limits the system performance improvement and the application to SM systems with an arbitrary number of antennas. To tackle this issue, the authors of \cite{Guo2016} and \cite{Guo2017} proposed an optimization strategy jointly considering the signal and spatial constellations for SM. \cite{Guo2017} showed that the joint optimization strategy achieves better performance than the sole signal constellation optimization strategies.  On the other hand, the gain is rather small yet costless, because the off-line design does not render any computational complexity for on-line data transmission.   \cite{Guo2017}  extended the optimization strategy regarding both signal and spatial constellations in GenSM systems using inter-channel interference (ICI)-free single-data-stream transmission. However, the design for more spectral-efficient multiple-data-stream GenSM was left unconsidered. Besides, little work is known on the constellation design  for GenQSM other than \cite{Iqbal2018} proposed a heuristic signal constellation optimization for QSM and \cite{Choi2018} studied a lattice-code-based constellation optimization strategy for GenQSM. These are based on  predefined codebooks and can be viewed as an optimization strategy in the discrete space.
 Generic optimization strategy in the continuous complex field regarding both signal and spatial constellations for GenQSM remains unexplored.
\subsubsection{Signal Shaping with CSIT}
With instantaneous/statistical CSIT, GenSM/GenQSM systems can benefit from adaptive signal shaping. For instance, adaptive signal constellation design was investigated for SM in \cite{Yang2012}; \cite{Rajashekar2013,Ntontin2013,Rajashekar2015} studied antenna selection for SM systems, which can be regarded as the adaptive spatial constellation optimization; A large body of literature such as \cite{Xiao2013,Guo2014,Zhang2016,Yang2015a,He2015,Lee2015,Yang2016,Zhang2017,Wang2015,Wang2016a,Wang2017,Wang2018,Cheng2018} probed into precoding schemes as well as their combinations with the adaptive signal or spatial constellation optimization.
However, most literature considered SM or single-data-stream GenSM. To the best of our knowledge,  few literature focused on the adaptive signal shaping for multiple-data-stream GenSM other than a recent work \cite{Cheng2018} which modifies a given multiple-dimensional signal constellation via a diagonal or  full precoder for each TAC.  However, such precoding-aided signal shaping methods in \cite{Cheng2018} are suboptimal. The reasons are twofolds: First, each TAC carries the same number of data symbol vectors in \cite{Cheng2018}, while due to the random nature of wireless channels, different TACs corresponding to separate channels have distinct information-carrying capabilities; Second, the data symbol vectors in the signal constellation of an activated TAC are modified by the same precoder, which can not guarantee the global signal shaping optimality, because the performance also highly depends on the previously given signal constellation. In addition, few work was dedicated to the adaptive shaping for GenQSM, except \cite{Li2017} which adjust the precoding process for QSM systems with a single IQ RF chain. 

In summary, both constellation design and adaptive signal shaping have been extensively investigated for SM and single-data-stream GenSM. A few of literature studied the signal shaping for QSM systems with a single IQ RF chain. However, the designs for multiple-data-stream GenSM/GenQSM with/without CSIT  call for a systematic investigation, which motivates us to fill this gap with the contributions listed infra. 
\subsection{Contributions}
\begin{itemize}
\item This paper formulates a unified signal shaping optimization problem for multiple-data-stream GenSM/GenQSM with/without CSIT, which aims to find the optimal transmit vector set under a size constraint, a unit average power constraint and a special sparsity constraint.
\item To solve the formulated problem, we reformulate the original problem to find the optimal signal constellations for each TAC. We optimize both the size and entry of all signal constellations. 
In particular, we suggest the use of a recursive design presented in \cite{Guo2017} for the size optimization. The entry optimization is formulated as a non-convex large-scale quadratically constrained quadratic programming (QCQP) problem, which can be solved by existing optimization techniques with   rather high computational complexity.
\item To reduce the computational complexity and also to facilitate the implementation in realistic systems, we adopt a codebook generated by using quadrature amplitude modulation (QAM) symbols to generate all feasible transmit vectors. Then, we propose a low-complexity progressive selection algorithm to choose the optimal transmit vector set. 
\item Numerical comparisons with existing designs are presented to verify the superiority of our designs. In particular, we compare the proposed joint constellation design with the sole signal constellation design in open-loop systems without CSIT and closed-loop systems with statistical CSIT, where the well-recognized best signal constellations (e.g., the binary phase shift keying, BPSK) are included in comparisons; we compare the proposed adaptive signal shaping methods with precoding-aided signal shaping methods in \cite{Cheng2018}. Comparison results show that the proposed optimization-based signal shaping (OBSS) approach considerably outperforms existing designs in literature and the proposed codebook-based signal shaping (CBSS) approach shows comparable performance with much lower complexity.  Moreover, the robustness of the proposed designs to channel uncertainty is also investigated by simulations.
\end{itemize}
\subsection{Organization}
The remainder of the paper is organized as follows. Section II describes the system model. Section III introduces the unified problem formulation for the signal shaping of multiple-data-stream GenSM/GenQSM with/without CSIT. In Sections IV and V, we introduce the OBSS and CBSS approaches, respectively. Numerical comparisons are presented in Section VI and  conclusions are drawn in Section VII.
\begin{figure*}[t]
  \centering
  \includegraphics[width=0.8\textwidth]{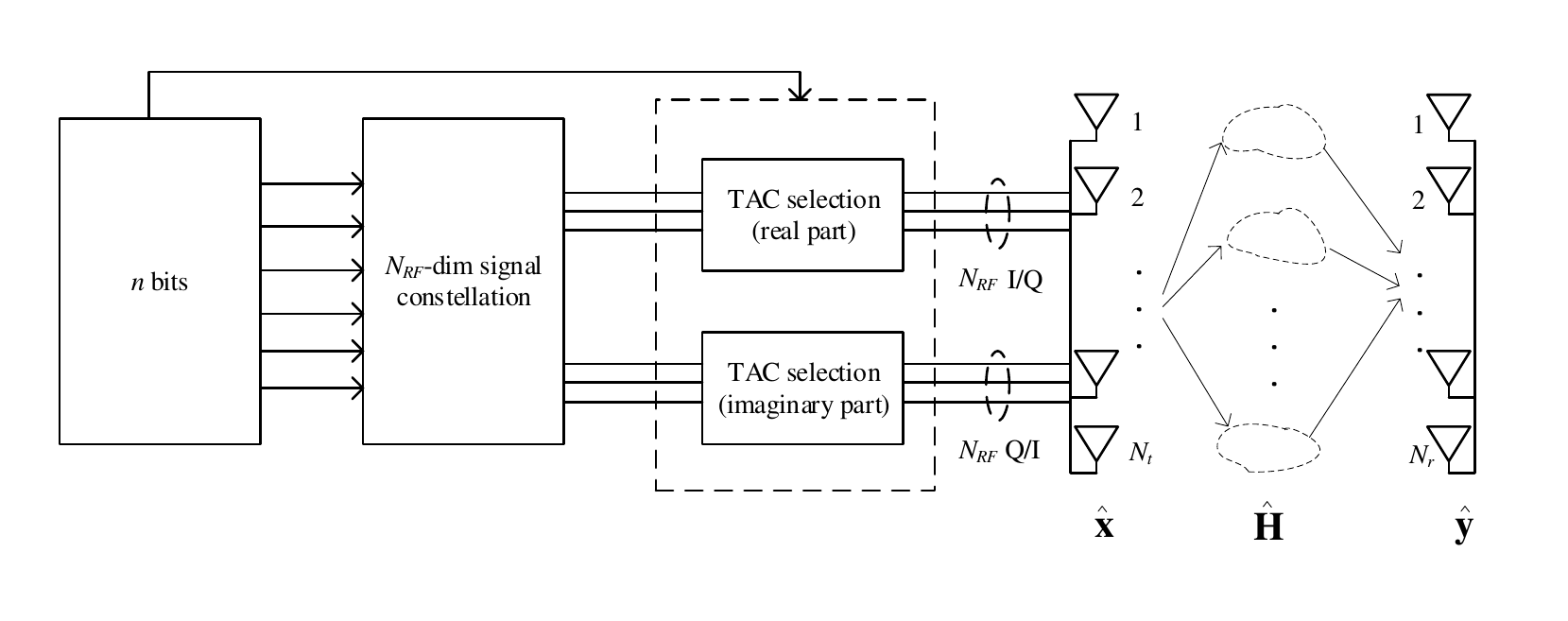}\\
 \caption{ A typical GenSM/GenQSM MIMO system.}
  \label{System_Model}
\end{figure*}

\subsection{Notations}
 In this paper, $a$ represents a scaler;  $\textbf{a}$ is a vector; $\textbf{A}$ stands for a matrix. 
$||\textbf{a}||_0$ and $||\textbf{a}||_2$ stand for the $l_0$ norm and the $l_2$ norm of $\textbf{a}$, receptively. $\diag{(\textbf{a})}$ is a diagonal matrix whose diagonal entries are from vector $\textbf{a}$. $\textbf{a}(i)$ denotes the $i$th entry of $\textbf{a}$. $\diag{\{\textbf{A}\}}$ stands for a vector formed by the diagonal elements of matrix $\textbf{A}$.
$\det(\textbf{A})$ represents the determination of matrix $\textbf{A}$. 
$\otimes$ stands for the Kronecker product. $(\cdot)^T$,  $(\cdot)^\dag$ stand for the transpose and the conjugate, respectively. $\mathbb{R}$ is the real domain; $\mathbb{Z}$ represents the integer domain; $\mathbb{C}$ stands for the complex domain. $\mathcal{CN}(\mu,\sigma^2)$ denotes the complex Gaussian distribution with  mean $\mu$ and variance $\sigma^2$. $\mathbb{E}(\cdot)$ represents the expectation operation. $\mathcal{A}$ is a set and $|\mathcal{A}|$ represents the size of  set $\mathcal{A}$. $[\textbf{A}]_{k,l}$ represents the $k$th row $l$th column entry of $\textbf{A}$. 
 $\Re{(\cdot)}$ and $\Im{(\cdot)}$ represent the real and imaginary parts, respectively.   $\left \lfloor{\cdot}\right \rfloor$ represents the floor operation. $\left(n\atop m\right)$ is a binomial coefficient. 

\section{System Model}
\subsection{System Framework}
In this paper, we consider an  ($N_t,N_r,N_{RF},n$) GenSM/GenQSM MIMO system as illustrated in Fig. \ref{System_Model}, where $N_t$ and $N_r$ represent the numbers of transmit and receive antennas; $N_{RF}$ is 
the number of RF chains and $n$ stands for the target transmission rate in bit per channel use (bpcu).  Unlike the conventional GenSM and GenQSM which map  data bits separately to the signal and spatial constellation points, we map $n$ data bits jointly to a transmit vector $\hat{\textbf{x}}\in\mathbb{C}^{N_t}$.
In such a system, the number of used TACs does not need to be a power of two, and the signal constellations used for different TACs are not required to be the same \cite{Guo2016}.

With $\hat{\textbf{x}}$ being transmitted, the receive signal vector $\hat{\textbf{y}}\in\mathbb{C}^{N_r}$  can be written as
\begin{equation}
\hat{\textbf{y}}=\sqrt{\rho}\hat{\textbf{H}}\hat{\textbf{x}}+\hat{\textbf{n}},
\end{equation}
where $\rho$ denotes the average receive SNR;  $\hat{\textbf{H}}\in\mathbb{C}^{N_r\times N_t}$ is the channel matrix  and $\hat{\textbf{n}}\in\mathbb{C}^{N_r}$ represents the complex Gaussian noise vector with zero mean and unit variance, i.e., $\hat{\textbf{n}}\sim\mathcal{CN}(\textbf{0},\textbf{I}_{N_r})$.
The transmission can be re-expressed in the real domain as
\begin{equation}
\textbf{y}=\sqrt{\rho}\textbf{H}\textbf{x}+\textbf{n},
\end{equation}
where 
\begin{equation}
\begin{split}
&\textbf{y}=
\begin{bmatrix}
\Re{(\hat{\textbf{y}})}\\
\Im{(\hat{\textbf{y}})}
\end{bmatrix}
,~\textbf{H}=
\begin{bmatrix}
\Re{(\hat{\textbf{H}})}&-\Im{(\hat{\textbf{H}})}\\
\Im{(\hat{\textbf{H}})}&\Re{(\hat{\textbf{H}})}
\end{bmatrix},\\
&\textbf{x}=
\begin{bmatrix}
\Re{(\hat{\textbf{x}})}\\
\Im{(\hat{\textbf{x}})}
\end{bmatrix},~\textbf{n}=
\begin{bmatrix}
\Re{(\hat{\textbf{n}})}\\
\Im{(\hat{\textbf{n}})}
\end{bmatrix}.
\end{split}
\end{equation}
We use $\mathcal{X}_N=\{\textbf{x}_{1},\textbf{x}_2,\cdots,\textbf{x}_{N}\}$ of size $N=2^n$ to represent the transmit vector set,  where  $\textbf{x}_i=[\Re(\hat{\textbf{x}}_i)^T,\Im(\hat{\textbf{x}}_i)^T]^T\in \mathbb{R}^{2N_t}$. It is assumed that the transmit vectors in $\mathcal{X}_N$ are under a unit average power constraint, which is expressed as
\begin{equation}
P(\mathcal{X}_N)=\mathbb{E}(||\textbf{x}||^2)=\frac{1}{N}\sum_{i=1}^{N}{\textbf{x}_i}^T{\textbf{x}_i}\leq 1.
\end{equation}
Besides, they are also under a sparsity constraint, because the number of  data streams sent via RF chains should be less than or equal to the number of RF chains. Specifically, for GenSM, the sparsity constraint is given by
\begin{equation}
||\textbf{x}_i^R+j\textbf{x}_i^I||_0\leq N_{RF},~i=1,2,\cdots,N,~ j=\sqrt{-1},
\end{equation}
while for GenQSM,  the sparsity constraint is given by
\begin{equation}
||\textbf{x}_i^R||_0\leq N_{RF},~
||\textbf{x}_i^I||_0\leq N_{RF},~i=1,2,\cdots,N.
\end{equation}
where $\textbf{x}_i^R\triangleq [\textbf{x}_i(1),\textbf{x}_i(2),\cdots,\textbf{x}_i(N_t)]^T=\Re{(\hat{\textbf{x}})}\in\mathbb{R}^{N_t}$ and $\textbf{x}_i^I\triangleq [\textbf{x}_i(N_t+1),\textbf{x}_i(N_t+2),\cdots,\textbf{x}_i(2N_t)]^T=\Im{(\hat{\textbf{x}})}\in\mathbb{R}^{N_t}$ represent the vectors being composed of the first $N_t$ entries and the last $N_t$ entries of vector $\textbf{x}_i$, respectively. 

\emph{Remark:} The sparsity constraints for the transmit vectors of GenSM and GenQSM are different from their conventional counterparts, because the positions of non-zero elements in the transmit vectors are constrained. That is, let $\mathcal{I}_i^R$ and $\mathcal{I}_i^I$ be the index sets of non-zero elements in $\textbf{x}_i^R$ and $\textbf{x}_i^I$, respectively. For GenSM, the sparsity constraint can be re-expressed as
\begin{equation}
|\mathcal{I}_i^I\cup\mathcal{I}_i^R|\leq N_{RF},~i=1,2,\cdots,N.
\end{equation}
For GenQSM, it is 
\begin{equation}
|\mathcal{I}_i^I|\leq N_{RF},~|\mathcal{I}_i^R|\leq N_{RF},~i=1,2,\cdots,N.
\end{equation}
For comparison purposes, we also give the conventional sparsity constraint without the position limitation as follows:
\begin{equation}
||\textbf{x}_i||_0\leq 2N_{RF},~i=1,2,\cdots,N,
\end{equation}
or 
\begin{equation}
|\mathcal{I}_i^I|+|\mathcal{I}_i^R|\leq 2N_{RF},~i=1,2,\cdots,N.
\end{equation}

\subsection{Channel Model}
In this paper, a specific transmit-correlated Rayleigh channel model is adopted  as \cite{Guo2017}
\begin{equation}
\hat{\textbf{H}}=\hat{\textbf{H}}_w\textbf{R}_{tx}^{1/2}.
\end{equation}
where $\hat{\textbf{H}}_w\in\mathbb{C}^{N_r\times N_t}$ represents a complex Gaussian matrix with $[\hat{\textbf{H}}_{w}]_{k,l}\sim\mathcal{CN}(0,1)$ and $\textbf{R}_{tx}\in\mathbb{C}^{N_t\times N_t}$ denotes the transmit correlation matrix.  The correlation weight matrix $\textbf{R}\in\mathbb{R}^{2N_t\times2N_t}$ can be written in the real domain as
\begin{equation}
\textbf{R}=\begin{bmatrix}
\Re{(\textbf{R}_{tx}^{1/2})}&-\Im{(\textbf{R}_{tx}^{1/2})}\\
\Im{(\textbf{R}_{tx}^{1/2})}&\Re{(\textbf{R}_{tx}^{1/2})}
\end{bmatrix}.
\end{equation}
Despite the specific channel model is used as an example for illustration purposes, it should be noted that the proposed optimization strategies and obtained results are applicable to generalized channel models.

\section{Unified Optimization Problem Formulation}
In this section, we formulate a unified optimization problem of the signal shaping for multiple-data-stream GenSM and GenQSM based on the MMED criterion in three cases that are without CSIT, and with statistical as well as instantaneous CSIT, respectively.
Without CSIT, $\mathcal{X}_N$ is designed for maximizing the minimum Euclidean distance of the transmit vectors by
\begin{equation}
\max d_{\min}(\mathcal{X}_N,\textbf{I}_{2N_t})=\max\min_{\textbf{x}_i\neq \textbf{x}_{i'}\in\mathcal{X}_N} ||\textbf{x}_i-\textbf{x}_{i'}||_2.
\end{equation}
With statistical CSIT (i.e., $\textbf{R}$), $\mathcal{X}_N$ is designed for maximizing the minimum Euclidean distance of the correlation wighted transmit vectors by
\begin{equation}
\max d_{\min}(\mathcal{X}_N,\textbf{R})=\max\min_{\textbf{x}_i\neq \textbf{x}_{i'}\in\mathcal{X}_N} ||\textbf{R}(\textbf{x}_i-\textbf{x}_{i'})||_2.
\end{equation}
With perfect CSIT, $\mathcal{X}_N$ is designed for maximizing the minimum Euclidean distance of noise-free receive signal vectors by
\begin{equation}
\max d_{\min}(\mathcal{X}_N,\textbf{H})=\max\min_{\textbf{x}_i\neq \textbf{x}_{i'}\in\mathcal{X}_N}||\textbf{H}(\textbf{x}_i-\textbf{x}_{i'})||_2.
\end{equation}
According to the optimization problems formulated above, we use a $2N_t$-column weight matrix $\textbf{A}$ to represent $\textbf{I}_{2N_t}$, $\textbf{R}$ or $\textbf{H}$, and rewrite three objective functions in a unified form  to be
\begin{equation}
\max d_{\min}(\mathcal{X}_N,\textbf{A})=\max\min_{\textbf{x}_i\neq \textbf{x}_{i'}\in\mathcal{X}_N}||\textbf{A}(\textbf{x}_i-\textbf{x}_{i'})||_2.
\end{equation}
Therefore, the signal shaping optimization problems for GenSM and GenQSM of the aforementioned three cases can be formulated as 
\begin{subequations}
\begin{align}
(\textbf{P1}):~~~~~\mathrm{Given}: &~\textbf{A}, N_{RF},N\notag\\
\mathrm{Find}:&~\mathcal{X}_N=\{\textbf{x}_{1},\textbf{x}_2,\cdots,\textbf{x}_{N}\}\notag\\
\mathrm{Maximize}:&~d_{\min}(\mathcal{X}_N,\textbf{A})\notag\\
\mathrm{Subject~to}:&~|\mathcal{X}_N|=N\\
&~P(\mathcal{X}_N)\leq 1 \\
&~||\textbf{x}_i^R+j\textbf{x}_i^I||_0\leq N_{RF}
\\&~i=1,2,\cdots, N,\notag
\end{align}
\end{subequations}
and 
\begin{subequations}
\begin{align}
(\textbf{P2}):~~~~\mathrm{Given}: &~\textbf{A}, N_{RF},N\notag\\
\mathrm{Find}:&~\mathcal{X}_N=\{\textbf{x}_{1},\textbf{x}_2,\cdots,\textbf{x}_{N}\}\notag\\
\mathrm{Maximize}:&~d_{\min}(\mathcal{X}_N,\textbf{A})\notag\\
\mathrm{Subject~to}:&~|\mathcal{X}_N|=N\\
&~P(\mathcal{X}_N)\leq 1\\
&~
||\textbf{x}_i^R||_0\leq N_{RF},~
||\textbf{x}_i^I||_0\leq N_{RF}
\\&~i=1,2,\cdots,N\notag,
\end{align}
\end{subequations}
respectively. In ($\textbf{P1}$) and ($\textbf{P2}$), (17a), (18a) represent the size constraints;  (17b), (18b) are the unit average power constraints and (17c), (18c) are the special sparsity constraints.

To replace the sparsity constraints in (17c) and (18c), we express $\textbf{x}_i$ as $\textbf{x}_i=\textbf{F}_k\textbf{s}_l^k$, where $\textbf{F}_k\in{\mathbb{R}^{2N_t\times 2N_{RF}}}$ is a matrix that corresponds to the $k$th TAC and $\textbf{s}_l^k\in{\mathbb{R}^{2N_{RF}}}$ is the $l$th data symbol vector when $\textbf{F}_k$ is activated. For GenSM, $\textbf{F}_k$ can be expressed as
\begin{equation}
\textbf{F}_k^{\mathrm{GenSM}}=\begin{bmatrix}
\textbf{C}_u&\textbf{0}\\
\textbf{0}&\textbf{C}_u
\end{bmatrix},
\end{equation}
where $\textbf{C}_u\in\mathbb{R}^{N_t\times N_{RF}}$ is an antenna selection matrix composed of $N_{RF}$ basis vectors of dimension $N_t$. 
Let $\mathcal{F}_{\mathrm{GenSM}}$  denote the set of all feasible  $\textbf{F}_k^{\mathrm{GenSM}}$.
Since there are totally $\left(N_{t}\atop N_{RF}\right)$ different $\textbf{C}_u$,  the number of feasible $\textbf{F}_k^{\mathrm{GenSM}}$  in $\mathcal{F}_{\mathrm{GenSM}}$ is $\left(N_{t}\atop N_{RF}\right)$, corresponding to  $\left(N_{t}\atop N_{RF}\right)$ feasible TACs. 
For GenQSM, $\textbf{F}_k$ can be given by
\begin{equation}
\textbf{F}_k^{\mathrm{GenQSM}}=\begin{bmatrix}
\textbf{C}_u&\textbf{0}\\
\textbf{0}&\textbf{C}_v
\end{bmatrix},
\end{equation}
where $\textbf{C}_u,\textbf{C}_v\in\mathbb{R}^{N_t\times N_{RF}}$ are two independent antenna selection matrices.  Let $\mathcal{F}_{\mathrm{GenQSM}}$ denote the set of all feasible  $\textbf{F}_k^{\mathrm{GenQSM}}$. As there are $\left(N_{t}\atop N_{RF}\right)$ different $\textbf{C}_u$ and $\textbf{C}_v$,  the number of feasible $\textbf{F}_k^{\mathrm{GenQSM}}$   in $\mathcal{F}_{\mathrm{GenQSM}}$ is $\left(N_{t}\atop N_{RF}\right)^2$, corresponding to $\left(N_{t}\atop N_{RF}\right)^2$ feasible TACs. 

 For unification, we use $\textbf{F}_k$ to represent  $\textbf{F}_k^{\mathrm{GenSM}}$  or $\textbf{F}_k^{\mathrm{GenQSM}}$ and $\mathcal{F}$ to represent $\mathcal{F}_{\mathrm{GenSM}}$ or $\mathcal{F}_{\mathrm{GenQSM}}$. The $k$th signal constellation  $\mathcal{S}_k$ is defined as the set of $\textbf{s}_l^k$ when $\textbf{F}_k$ is activated and the set of all signal constellations is represented by $\mathcal{Z}=\left\{\mathcal{S}_1,\mathcal{S}_2,\cdots,\mathcal{S}_{|\mathcal{F}|}\right\}$. Based on these denotations,  $(\textbf{P1})$ and $(\textbf{P2})$ can be expressed in a unified manner as follows:
\begin{equation}
\begin{split}
(\textbf{OP}):~~~~~\mathrm{Given}: &~\textbf{A},N,\mathcal{F}\\
\mathrm{Find}:&~\mathcal{Z}=\left\{\mathcal{S}_1,\mathcal{S}_2,\cdots,\mathcal{S}_{|\mathcal{F}|}\right\}\\
\mathrm{Maximize}:&~d_{\min}(\mathcal{F},\mathcal{Z},\textbf{A})\\
\mathrm{Subject~to}:&~\sum_{k=1}^{|\mathcal{F}|}|\mathcal{S}_k|=N\\
&~P(\mathcal{Z})\leq 1, \\
\end{split}
\end{equation}
where 
\begin{equation}
d_{\min}(\mathcal{F},\mathcal{Z},\textbf{A})=\min_{{{\textbf{F}_k\textbf{s}_l^k\neq \textbf{F}_{k'}\textbf{s}_{l'}^{k'}\atop \textbf{F}_k,\textbf{F}_{k'}\in\mathcal{F}}\atop\textbf{s}_l^k\in\mathcal{S}_k,~\textbf{s}_{l'}^{k'}\in\mathcal{S}_{k'} }\atop\mathcal{S}_k,\mathcal{S}_{k'}\in\mathcal{Z}}||\textbf{H}(\textbf{F}_k\textbf{s}_l^k-\textbf{F}_{k'}\textbf{s}_{l'}^{k'})||_2,
\end{equation}
and 
\begin{equation}
\begin{split}
P(\mathcal{Z})
=\frac{1}{N}\sum_{k=1}^{|\mathcal{Z}|}\sum_{l=1}^{|\mathcal{S}_k|}(\textbf{s}_l^k)^T{\textbf{s}_l^k}.
\end{split}
\end{equation}

\section{ Optimization-Based Signal Shaping}
Since the problem (\textbf{OP}) for optimizing $\mathcal{X}_N$ has been reformulated to search the optimal signal constellations $\mathcal{S}_1$, $\mathcal{S}_2$,$\dots$,$\mathcal{S}_{|\mathcal{F}|}$ in the last section, the problem becomes a set optimization problem including the set size optimization and the set entry optimization. We analyze both sub-problems with details in the following subsections.

\subsection{Set Size Optimization}
The set size optimization is a non-negative integer programming satisfying $\sum_{k=1}^{|\mathcal{F}|}|\mathcal{S}_k|=N$. There are a total number of $\left(N+|\mathcal{F}|-1\atop|\mathcal{F}|-1\right)$ feasible solutions \cite{Sheldon2002}, which is rather large. Taking a MIMO system with $N_t=4$, $N_{RF}=2$ and $N=16$ (i.e., $n=4$ bpcu) as an example, $|\mathcal{F}_{\mathrm{GenSM}}|=\left(4\atop 2\right)=6$ and $|\mathcal{F}_{\mathrm{GenQSM}}|=\left(4\atop 2\right)^2=36$.  We can easily calculate that there are $\left(16+6-1\atop 6-1\right)\approx2\times 10^4 $ feasible size solutions for GenSM and $\left(16+36-1\atop 36-1\right)\approx7.2\times10^{12}$ solutions for GenQSM, respectively. Therefore,  exhaustive search for the optimal solution is undesirable, because set entry optimization is needed for each feasible set size solution. In \cite{Guo2017} which considers single-data-stream transmission without CSIT,  a greedy recursive method was proposed by finding the optimal set size solution for $\mathcal{X}_N$ with size $N$ according to $\mathcal{X}_{N-1}$ with size $N-1$. The recursive method shows comparable performance to the exhaustive search  in the single-data-steam cases but demands much lower complexity \cite{Guo2017}. Thus, we also suggest the use of its extension in multiple-data-steam cases. To introduce the extension, we define a 
constellation partition matrix 
$\textbf{W}_N\in\mathbb{R}^{2 N_t\times 2NN_{RF}}$ for $\mathcal{X}_N$ with size $N$ as
\begin{equation}
\textbf{W}_N\triangleq\left[\overbrace{\textbf{F}_1,\cdots,\textbf{F}_1}^{|\mathcal{S}_1|},\overbrace{\textbf{F}_2,\cdots,\textbf{F}_2}^{|\mathcal{S}_2|},\cdots,\cdots,\overbrace{\mathcal{\textbf{F}_{|\mathcal{F}|},\cdots,\textbf{F}_{|\mathcal{F}|}}}^{|\mathcal{S}_{|\mathcal{F}|}|}\right],
\end{equation}
which can be regarded as an indicator of the signal constellation sizes $\{|\mathcal{S}_k|\}$ and the number of $\textbf{F}_k$ in $\textbf{W}_N$ indicates the signal constellation size $|\mathcal{S}_k|$ for the $k$th TAC. With the definition of $\textbf{W}_N$, the extension of the recursive design in multiple-data-stream cases can be described as follows. Given $\textbf{W}_{N-1}$, we can choose an $\textbf{F}_k\in \mathcal{F}$ to adjoin $\textbf{W}_{N-1}$ generating $|\mathcal{F}|$ candidates of $\textbf{W}_{N}$. For each candidate of $\textbf{W}_{N}$, we perform set size optimization and obtain the corresponding candidates of $\mathcal{X}_N$. Then, by comparing all the  candidates of $\mathcal{X}_N$, we can obtain the optimal $\mathcal{X}_N$ among all the candidates and the corresponding optimal $\textbf{W}_{N}$.  Based on this principle, we use the optimal $\mathcal{X}_{2}$ and $\textbf{W}_{2}$, which can be obtained by exhaustive search, to find the optimal $\mathcal{X}_{3}$ and $\textbf{W}_{3}$, then $\mathcal{X}_{4}$ and $\textbf{W}_{4}$ and so on until the size constraint is satisfied.

\subsection{Set Entry Optimization}
Given a fixed $\textbf{W}_N$, the sizes of $\mathcal{S}_1,\mathcal{S}_2,\cdots,\mathcal{S}_{|\mathcal{F}|}$ are determined and we now need to optimize the set entries in each set to maximize $d_{\min}(\mathcal{F},\mathcal{Z},\textbf{A})$ in (\textbf{OP}). To solve the problem,
we define $\textbf{S}_l^k\triangleq\diag(\textbf{s}_l^k)$ for all $k=1,2,\cdots,|\mathcal{F}|$, $l=1,2,\cdots,|\mathcal{S}_k|$, and a diagonal matrix $\textbf{D}_{\textbf{q}}$ of dimension ${2NN_{RF}\times 2NN_{RF}}$ as
\begin{equation}
\textbf{D}_{\textbf{q}}\triangleq 
\begin{bmatrix}
    ~\textbf{S}_1^1 & \textbf{0} & \textbf{0}&\cdots&\textbf{0}&\textbf{0}&  \textbf{0}~ \\
    ~\textbf{0} & \ddots& \textbf{0}& \cdots&\cdot&\textbf{0}& \textbf{0}~ \\
     ~\textbf{0} & \textbf{0} & \textbf{S}_{|\mathcal{S}_1|}^1&  \cdots&\textbf{0}&\cdot&\textbf{0}~\\
~\vdots& \vdots & \vdots& \ddots& \vdots&\vdots&\vdots~\\
    ~\textbf{0} & \cdot & \textbf{0} &\cdots &\textbf{S}_{1}^{|\mathcal{F}|}& \textbf{0}&  \textbf{0}~\\
 ~\textbf{0} & \textbf{0} & \cdot &\cdots &\textbf{0}& \ddots&  \textbf{0}~\\
 ~\textbf{0} & \textbf{0} & \textbf{0} &\cdots &\textbf{0}& \textbf{0}&  \textbf{S}_{|\mathcal{S}_{|\mathcal{F}|}|}^{|\mathcal{F}|}~
\end{bmatrix},
\end{equation}
as well as a vector $\textbf{e}_i\in\mathbb{R}^{2NN_{RF}}$ as $\textbf{e}_i\triangleq\textbf{g}_i\otimes \textbf{1}_{2N_{RF}}$ where $\textbf{g}_i$ is the $i$th $N$-dimensional vector basis with all zeros expect the $i$th entry being one. Based on these definitions, the square of the pairwise Euclidean distances can be expressed as
\begin{equation}\label{eqD}
\begin{split}
||\textbf{A}\textbf{x}_i-\textbf{A}\textbf{x}_{i'}||_2^2&=||\textbf{A}\textbf{W}_N\textbf{D}_{\textbf{q}}\textbf{e}_i-\textbf{A}\textbf{W}_N\textbf{D}_{\textbf{q}}\textbf{e}_{i'}||_2^2\\
&=(\textbf{e}_i-\textbf{e}_{i'})^T\textbf{D}_\textbf{q}^T\textbf{W}_N^T\textbf{A}^T\textbf{A}\textbf{W}_N\textbf{D}_\textbf{q}(\textbf{e}_i-\textbf{e}_{i'})\\
&=\tr\left(\textbf{D}_\textbf{q}^T\textbf{R}_{\textbf{AW}}\textbf{D}_\textbf{q}\Delta\textbf{E}_{i{i'}}\right),
\end{split}
\end{equation}
where $\textbf{R}_{\textbf{A}\textbf{W}}=\textbf{W}_N^T\textbf{A}^T\textbf{A}\textbf{W}_N$ and $\Delta\textbf{E}_{i{i'}}=(\textbf{e}_i-\textbf{e}_{i'})(\textbf{e}_i-\textbf{e}_{i'})^T$.
Adopting the equality $\tr(\textbf{D}_{\textbf{u}}\textbf{U}\textbf{D}_{\textbf{v}}\textbf{V}^T)=\textbf{u}^T(\textbf{U}\odot\textbf{V})\textbf{v}$, where $\textbf{D}_{\textbf{u}}=\diag(\textbf{u})$ and $\textbf{D}_{\textbf{v}}=\diag(\textbf{v})$, we rewrite (\ref{eqD}) as
\begin{equation}
||\textbf{A}\textbf{x}_i-\textbf{A}\textbf{x}_{i'}||_2^2=\textbf{q}^T\textbf{Q}_{i{i'}}\textbf{q},
\end{equation}
where $\textbf{q}= \diag\{\textbf{D}_\textbf{q}\}\in\mathbb{R}^{2NN_{RF}}$ and $\textbf{Q}_{i{i'}}=\textbf{R}_{\textbf{A}\textbf{W}}\odot\Delta\textbf{E}_{i{i'}}^T\in\mathbb{R}^{2NN_{RF}\times2NN_{RF}}$.
As a consequence, the unit power constraint can be expressed as
\begin{equation}
P(\mathcal{Z})
=\frac{1}{N}\tr\left(\textbf{D}_\textbf{q}\textbf{D}_\textbf{q}^T\right)
=\frac{1}{N}\textbf{q}^T\textbf{q}\leq 1.
\end{equation}
Based on the above reformulations, the set entry optimization  becomes
\begin{equation}
\begin{split}
(\textbf{S-OP}):~~\mathrm{Given}: &~\textbf{Q}_{ii'}, \forall i\neq i'\in\{1,2,\cdots,N\}\\
\mathrm{Find}:&~\textbf{q}\\
\mathrm{Maximize}:&~\min \textbf{q}^T\textbf{Q}_{ij}\textbf{q}\\
\mathrm{Subject~to}:&~\textbf{q}^T\textbf{q}\leq N
\end{split}
\end{equation}
By introducing an auxiliary variable $\tau$, the optimization problem can be equivalently transformed to be
\begin{equation}
\begin{split}
(\textbf{S-OP-a}):~\mathrm{Given}: &~\textbf{Q}_{ii'}, \forall i\neq i'\in\{1,2,\cdots,N\}\\
\mathrm{Find}:&~\textbf{q},\tau\\
\mathrm{Maximize}:&~\tau\\
\mathrm{Subject~to}:&~\textbf{q}^T\textbf{Q}_{ij}\textbf{q}\geq \tau,\forall i\neq i'\in\{1,2,\cdots,N\}\\
&~\textbf{q}^T\textbf{q}\leq N
\end{split}
\end{equation}
Problem (\textbf{S-OP-a}) is a non-convex large-scale QCQP problem  with $2NN_{RF}$ variables and $\left(N\atop 2\right)$ constraints, and  can be solved by the iterative algorithm developed in  \cite{Lee2015} with complexity about $\mathcal{O}(N^4N_{RF}^2)$ in each iteration\footnote{We omit the other terms in the complexity analysis in \cite{Lee2015} and \cite{Cheng2018} for simplicity, since $N$ is much larger than other parameters.}. 
Moreover, problem (\textbf{S-OP-a}) is equivalent to the following problem:
\begin{equation}
\begin{split}
(\textbf{S-OP-b}):~\mathrm{Given}: &~\textbf{Q}_{ii'}, \forall i\neq i'\in\{1,2,\cdots,N\}\\
\mathrm{Find}:&~\textbf{q}\\
\mathrm{Maximize}:&~\textbf{q}^T\textbf{q}\\
\mathrm{Subject~to}:&~\textbf{q}^T\textbf{Q}_{ij}\textbf{q}\geq d,\forall i\neq i'\in\{1,2,\cdots,N\},
\end{split}
\end{equation}
where $d$ is a target minimum distance. The problem (\textbf{S-OP-b}) can be solved by the Lagrangian method  developed in \cite{Cheng2018}. The complexity is also  around $\mathcal{O}(N^4N_{RF}^2)$ in each iteration$^1$ but fortunately the algorithm in  \cite{Cheng2018} converges faster than that in \cite{Lee2015}.

\begin{algorithm}[t]
\caption{OBSS Procedure}
\label{alg:A}
\begin{algorithmic}
\STATE  \textbf{Input:} $\textbf{A}$, $N_t$, $N_{RF}$, $\mathcal{F}$ and $N$ 
\STATE  \textbf{Output:} $\mathcal{X}_N$
\STATE {\%\% \emph{Exhaustive Search for $\mathcal{X}_2$ and $\textbf{W}_2$}\\Generate $|\mathcal{F}|^2$ feasible candidates of $\textbf{W}_2$.\\Perform set entry optimization by solving (\textbf{S-OP}) to obtain  the $|\mathcal{F}|^2$ candidates of $\mathcal{X}_2$.\\Compare all  the candidates of $\mathcal{X}_2$ in terms of minimum Euclidean distance to find the optimal one and the corresponding  $\textbf{W}_2$.}
\STATE {\%\% \emph{Initialization}\\Initialize $t=3$.}
\STATE {\%\% \emph{Recursive Optimization for $\mathcal{X}_{N}$ and $\textbf{W}_N$}}
\REPEAT
\STATE{ Generate $|\mathcal{X}|$ feasible $\textbf{W}_t$ based on $\textbf{W}_{t-1}$. }
\STATE{Perform set entry optimization by solving (\textbf{S-OP}) to obtain  the $|\mathcal{F}|$ candidates for $\mathcal{X}_t$.\\Compare all the candidates of $\mathcal{X}_t$ in terms of minimum Euclidean distance  to find the optimal one and the related $\textbf{W}_t$.}
\STATE{Update} $t\leftarrow t+1$.
\UNTIL{$t>N$.}
\STATE{Output the optimized $\mathcal{X}_N$.} 
\end{algorithmic}
\end{algorithm}

\subsection{Complexity Analysis}
For clearly viewing the detailed design procedure of OBSS, we list it in Algorithm 1. The complexity of Algorithm 1 is determined by the exhaustive search for $\mathcal{X}_2$ and the recursive optimization for $\mathcal{X}_N$. In the exhaustive search for $\mathcal{X}_2$, we need to solve the (\textbf{S-OP}) problem for $|F|^2$ times. The aggregate complexity is $\mathcal{O}(|\mathcal{F}|^2N_{\mathrm{iter}_2}2^4N_{RF}^2)$, where $N_{\mathrm{iter}_2}$ denotes the number of iterations for solving problem (\textbf{S-OP}). It is obvious that the complexity is rather low and thereby negligible. In the recursive optimization for $\mathcal{X}_N$, we need to solve the (\textbf{S-OP}) problem for $|F|$ times for designing $\mathcal{X}_t$ and the aggregate complexity is
\begin{equation}
\mathrm{C}=\sum_{t=3}^{N}\mathcal{O}\left(N_{\mathrm{iter}_t}|\mathcal{F}|t^4N_{RF}^2\right)\approx \mathcal{O}(N_{\mathrm{iter}_t}|\mathcal{F}|N^5N_{RF}^2),
\end{equation}
where $N_{\mathrm{iter}_t}$ represents the number of iterations for solving problem (\textbf{S-OP}) to attain $\mathcal{X}_t$ and we assume $\{N_{\mathrm{iter}_t}\}$ are of the same order for any $t$.
\begin{table*}[t]
\centering
\caption{Comparisons Among Different Signal Shaping Approaches}\label{tab1}
    \begin{tabular}{ | c | c |c|c|c|}
    \hline
Approaches&Number of variables &Number of constraints&Complexity&Applications\\
\hline
OBSS & $2NN_{RF}$&$\left(N\atop2\right)$&$ \mathcal{O}(N_{\mathrm{iter}_t}|\mathcal{F}|N^5N_{RF}^2)$&GenSM/GenQSM\\
\hline
Diagonal precoding \cite{Cheng2018} & $2|\mathcal{F}_s|N_{RF}$&$\left(N\atop2\right)$&$\mathcal{O}(N_{\mathrm{iter}}^D|\mathcal{F}_s|^2N_{RF}^2N^2)$&GenSM\\
\hline
Full precoding \cite{Cheng2018} & $2|\mathcal{F}_s|N_{RF}^2$& $\left(N\atop2\right)$ &$\mathcal{O}(N_{\mathrm{iter}}^F|\mathcal{F}_s|^2N_{RF}^4N^2)$&GenSM\\
\hline
CBSS&-&-&$\mathcal{O}(N_c^2)$&GenSM/GenQSM
\\
\hline
    \end{tabular}
\end{table*}
\subsection{Remarks}
The proposed OBSS approach aims to directly optimize the transmit vectors, which is different from the precoding-aided signal shaping in \cite{Cheng2018} that modifies a given signal constellation of size $M^{N_{RF}}$ via a diagonal preceding matrix or a full precoding matrix for each TAC, where they assume $M$-ary modulation is adopted for each data stream transmission. Even though the diagonal precoder and the full precoder can be designed by solving a similar QCQP problem as (\textbf{S-OP}), there exists a difference in the number of optimization variables. For the diagonal precoder optimization, the total number of variables is $2|\mathcal{F}_s|N_{RF}$, while for the full precoder optimization, the total number of variables is $2|\mathcal{F}_s|N_{RF}^2$, where $\mathcal{F}_s$ is a selected subset of $\mathcal{F}$ and $|\mathcal{F}_s|=2^{\left \lfloor{\log_2 |\mathcal{F}|}\right \rfloor}$.
 Compared with  $2|\mathcal{F}_s|N_{RF}$ and $2|\mathcal{F}_s|N_{RF}^2$, the number of variables $2NN_{RF}$ in (\textbf{S-OP}) is much larger,  because $N=|\mathcal{F}_s|M^{N_{RF}}$. Additionally, the number of constraints in (\textbf{S-OP}) and that in the precoding-aided signal shaping in \cite{Cheng2018} are the same. That is, the scale of the QCQP problem in the proposed OBSS approach is larger than those formulated in precoding-aided shaping methods in \cite{Cheng2018}. 
Anyway, the proposed OBSS approach can be treated as a generalized design and hence can yield better performance.
Also, the global optimality of the proposed OBSS approach can not be guaranteed, because of the greedy set size optimization and the non-convexity of (\textbf{S-OP}) in the set entry optimization.
In summary, the OBSS approach is proposed for the sake of performance enhancement, whereas its complexity of which is rather high. It is affordable for off-line designs without CSIT or on-line designs with long-term invariant statistical CSIT. However, the complexity is a heavy burden for large-size signal shaping designs or adaptive designs with instantaneous CSIT. For adaptive designs, the complexity needs to be greatly reduced.

\section{Codebook-Based Signal Shaping}
To alleviate the heavy computation burden of OBSS, we propose an alternative low-complexity CBSS approach. We use a codebook generated by $M_c$-ary QAM modulation symbols, which is inspired by the work in \cite{Choi2018}. With the codebook, there are $M_c^{N_{RF}}$ feasible signal constellation points and a total number of $N_c=|\mathcal{F}|M_c^{N_{RF}}$ feasible transmit vectors in the candidate set $\mathcal{X}_c$. Then, we need to select a subset of size $N$. To reduce the selection complexity, we progressively select vectors to maximize the 
constellation figure merit  (CFM), which is equivalent to the MMED criterion under normalized power constraint and defined by \cite{Wang2016}
\begin{equation}
\mathrm{CFM}(\mathcal{X}_N)\triangleq\frac{ d_{\min}(\mathcal{X}_N,\textbf{A})^2}{P(\mathcal{X}_N)}.
\end{equation}
This selection procedure is illustrated in Algorithm 2. The complexity of Algorithm 2 is dominated by the calculations of pairwise Euclidean distances, which is of the order $\mathcal{O}(N_c^2)$. For clarity, comprehensive comparisons among the proposed approach and other existing approaches are given in Table I. From Table I, we find that its complexity is much less than existing approaches since it does not need to solve the large-scale QCQP problems, which enables its extensions to large-size signal shaping designs or adaptive designs with instantaneous CSIT.
\begin{algorithm}[htb] 
\caption{Progressive Selection Algorithm}
\label{alg:greedy}
\begin{algorithmic}
\STATE  \textbf{Input:} $\textbf{A}$, $N_c$, $N$ and $\mathcal{X}_c$
\STATE  \textbf{Output:} $\mathcal{X}$
\STATE Compute the powers of $\forall \textbf{x}_i\in\mathcal{X}_c$ and $||\textbf{A}(\textbf{x}_i-\textbf{x}_{i'})||$, where $\textbf{x}_{i}, \neq\textbf{x}_{i'}\in\mathcal{X}_c$. Save them for the computation of the CFMs.
\STATE {\%\% \emph{Exhaustive Search for $\mathcal{X}_2$}\\Generate $\left(N_c\atop 2\right)$ feasible candidates of  $\mathcal{X}_2$.
\\Compare all candidates in terms of CFM to find the optimal one.}
\STATE {\%\% \emph{Initialization}\\Initialize $t=3$.}
\STATE \%\% \emph{Progressive Selection to Design $\mathcal{X}_N$}
\REPEAT
\STATE{ Select a vector $\textbf{x}_i\in\mathcal{X}_c\setminus\mathcal{X}_{t-1}$ and add it into $\mathcal{X}_{t-1}$ as the candidates of $\mathcal{X}_t$.}
\STATE{Compare all candidates of $\mathcal{X}_t$ in terms of CFM  and find the optimal one.}
\STATE{Update} $t\leftarrow t+1$.
\UNTIL{$t>N$.}
\STATE{Output the selected $\mathcal{X}_N$.} 
\end{algorithmic}
 \end{algorithm}

\section{Simulations and Discussions}
In the simulations, we investigate the performance of the proposed OBSS and CBSS approaches in variously configured $(N_t,N_r,N_{RF},n)$ GenSM and GenQSM MIMO systems. The transmit-correlated Rayleigh channel model is adopted as described in Section II-B and the correlation matrix is defined by \cite{Guo2017}
\begin{equation}
\begin{split}
[\textbf{R}_{tx}]_{k,l}=\left\{
\begin{array}{rcl}
\delta^{k-l},    &     k\leq l\\
 (\delta^{l-k})^{\dag},   &   l>k
\end{array} \right. 
\end{split}
k,l=1,\cdots,N_t
\end{equation}
where $\delta$ represents the transmit correlation coefficient. Moreover, it is assumed that the transmitter and receiver both know $\delta$. We compare different signal shaping methods not only in the minimum Euclidean distance but also in the symbol error rate (SER), because the SER as one of the most important metrics of communication systems is determined by the transmit vector set $\mathcal{X}_N$. 
Over transmit-correlated Raleigh model, the asymptotic upper bound of the SER given $\mathcal{X}_N$ can be expressed as \cite{Guo2017}
\begin{equation}\label{UB}
\overline{P_s}=c\sum_{i=1}^{N}\sum_{i'=1,i\neq i'}^{N}||\textbf{R}(\textbf{x}_i-\textbf{x}_i')||_2^{-2N_{r}},
\end{equation}
where $c=\frac{\rho^{-N_r}}{N}\left(2N_r-1\atop N_r\right)$.

For clarity, we divide the section into three subsections. 
 In the first subsection, we compare the proposed OBSS solution with the sole signal constellation optimized results in open-loop systems without CSIT and closed-loop systems with statistical CSIT, where the optimal signal constellation has already been known and adopted for comparison.
In the second subsection, we investigate the proposed OBSS and CBSS optimization strategies in closed-loop systems with instantaneous CSIT and compare them with the precoding-aided signal shaping methods in \cite{Cheng2018}. 
In the last subsection, we investigate the impact of channel uncertainty on the performance to show the robustness of the proposed designs.

\subsection{Superiority of Proposed Design in Open-Loop Systems without CSIT and Closed-Loop Systems with Statistical CSIT}
\begin{table}[t]
\centering
\caption{Minimum Euclidean Distance Comparisons in $(3,2,2,3)$ GenSM MIMO Systems}\label{tab2}
    \begin{tabular}{ | c | c|c|c|}
    \hline
-&$\delta=0$&$\delta=0.1$&$\delta=0.3$\\
\hline
BPSK&$1$&$0.7007$&$0.4850$\\
\hline
OBSS&$1.4768$&$1.5738$&$1.6508$\\
\hline
    \end{tabular}
\end{table}

\begin{figure}[t]
  \centering
  \includegraphics[width=0.5\textwidth]{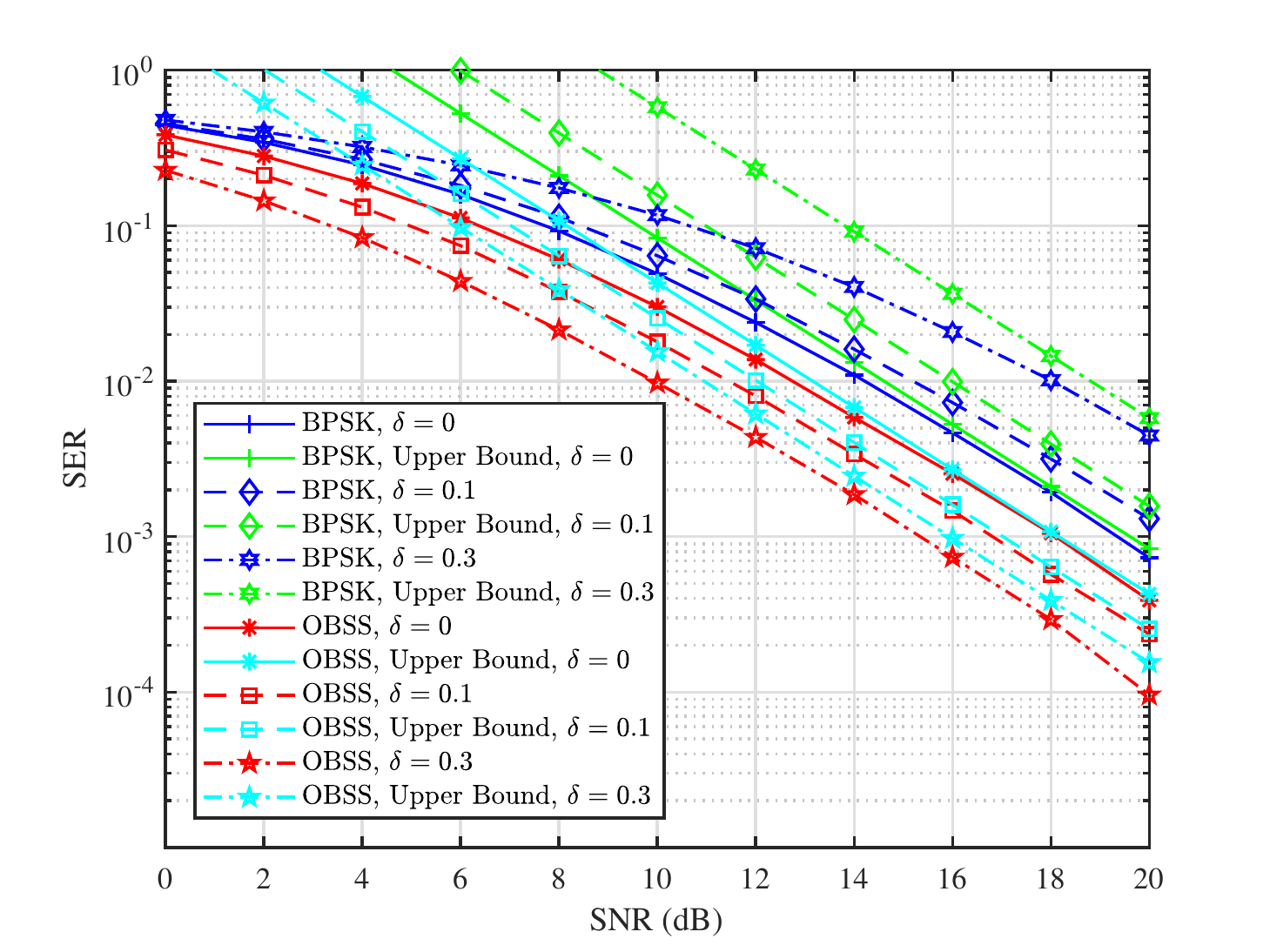}\\
 \caption{SER comparisons in $(3,2,2,3)$ GenSM MIMO systems.}
  \label{result1a}
\end{figure}

Firstly, we compare the proposed OBSS design with the well-recognized optimal signal constellation in $(3,2,2,3)$  GenSM MIMO systems under different channel conditions.  In $(3,2,2,3)$ GenSM systems, two TACs selected from all $\left(3\atop 2\right)=3$ ones can provide a rate of $1$ bpcu, subtracting which we can derive that binary modulation is used for $2$-data-streams carrying $2$ bpcu. The BPSK as the optimal binary signal constellation is adopted for comparison. By the proposed OBSS approach, we list the designed transmit vectors in Appendix A when $\delta=0$, $0.1$ and $0.3$. It should be noted that the optimal transmit vectors that achieve the maximized Euclidean distances are not unique and the given transmit vectors are just a specific realization.  
We compare them with BPSK-based shaping in terms of the minimum Euclidean distances $d_{\min}(\mathcal{X}_8,\textbf{R})$ as listed in Table II. It is found that the proposed OBSS optimization strategy has a much larger $d_{\min}(\mathcal{X}_8,\textbf{R})$ for $\delta=0$, $\delta=0.1$ and $\delta=0.3$ and as $\delta$ increases, $d_{\min}(\mathcal{X}_8,\textbf{R})$ also increases. 
The SER comparisons are  illustrated in Fig. \ref{result1a}, where the analytical upper bounds in (\ref{UB}) are also included. It demonstrates that OBSS outperforms $(3,2,2,3)$ GenSM with BPSK by $1$ dB over independent Rayleigh channels (i.e., $\delta=0$). The gain is achieved at no expense. Over the transmit-correlated channels,  results show that the OBSS approach can bring a higher gain. Specifically, more than $3.5$ dB and $8$ dB are achieved at an SER of $10^{-2}$ when $\delta=0.1$ and $0.3$ respectively owing to the joint optimization of spatial and signal constellation.

\begin{figure}[t]
  \centering
  \includegraphics[width=0.5\textwidth]{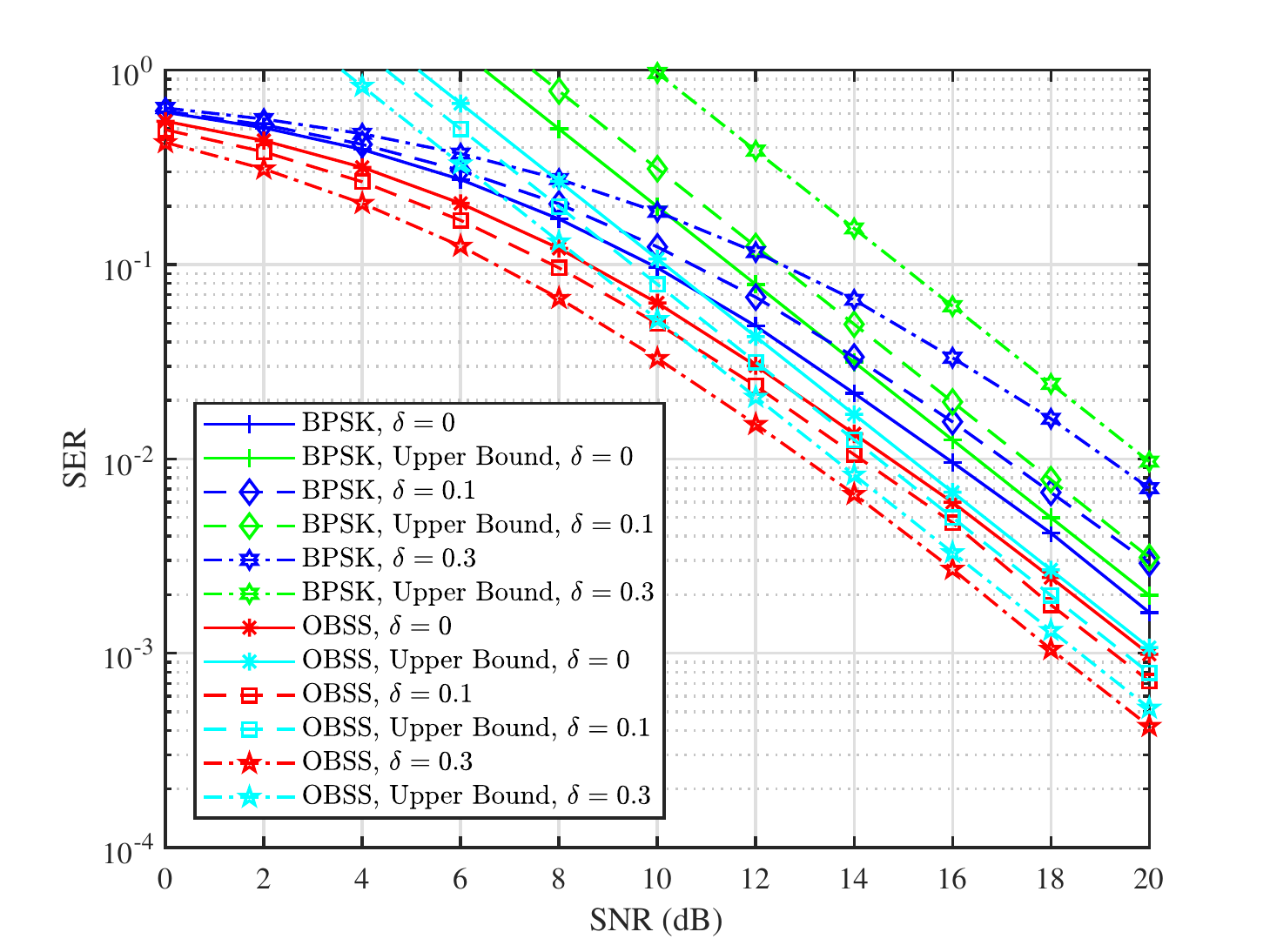}\\
 \caption{SER comparisons in $(4,2,2,4)$ GenSM MIMO systems.}
  \label{result1b}
\end{figure}

 To show more results, we also make the comparisons regarding SER in $(4,2,2,4)$ GenSM MIMO systems where BPSK is also employed for comparison. Fig. \ref{result1b} demonstrates similar trends that more than $1$ dB, $3$ dB and $6$ dB are achieved compared to GenSM with BPSK when $\delta=0$, $0.1$ and $0.3$, respectively. All above comparisons show that OBSS can be used to combat transmit correlation and above all, it can benefit from the transmit correlation. The price paid to achieve such substantial gains is just the knowledge of $\delta$. Knowing $\delta$ is sometimes costless as $\delta$ is simply determined by the antenna deployment.

\begin{figure}[t]
  \centering
  \includegraphics[width=0.5\textwidth]{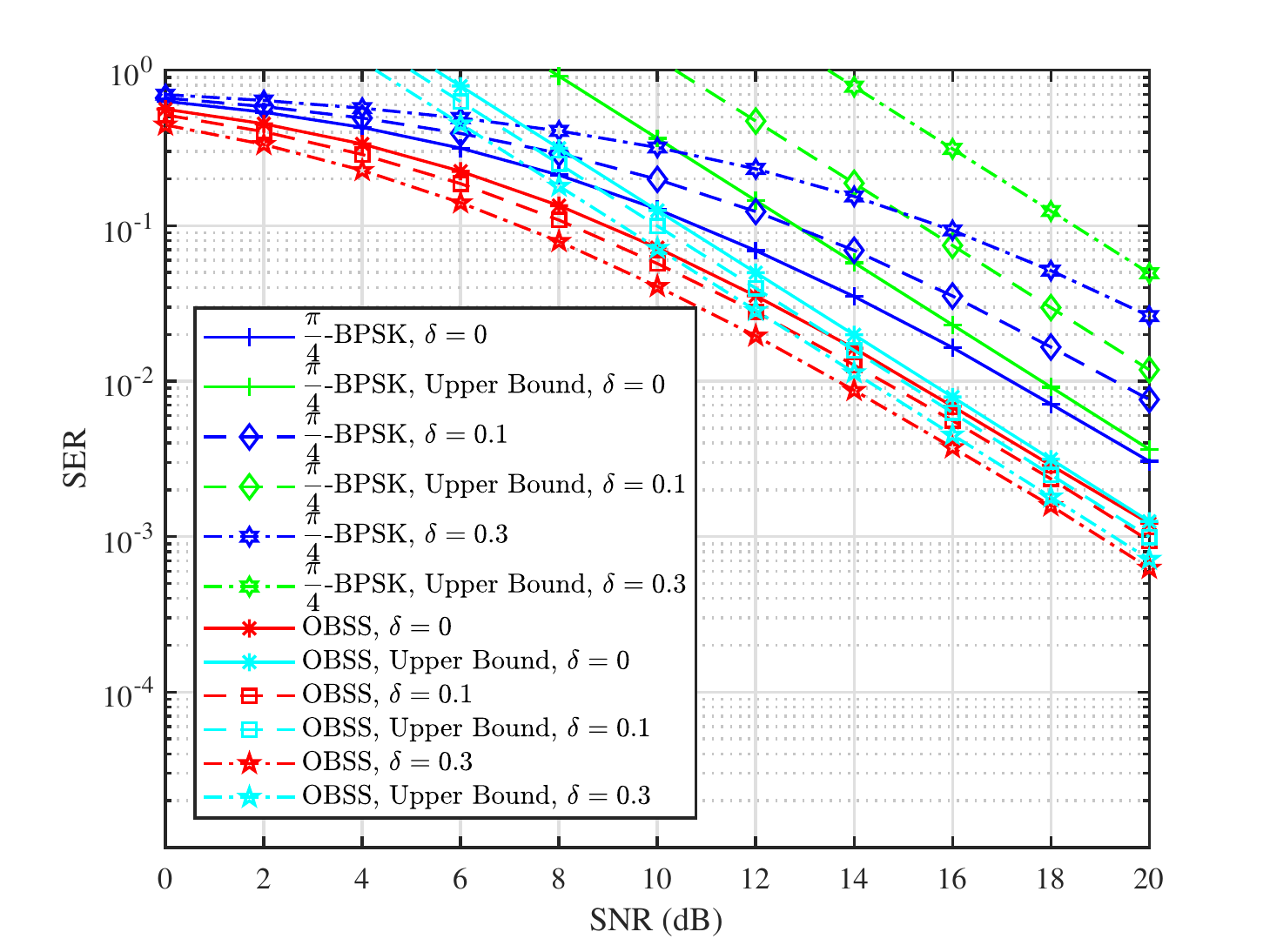}\\
 \caption{SER comparisons in $(3,2,2,4)$ GenQSM MIMO systems.}
  \label{result1c}
\end{figure}

Secondly, to validate the performance superiority of the proposed optimization strategies in GenQSM systems, we make the minimum Euclidean distance and SER comparisons in $(3,2,2,4)$ GenQSM MIMO systems. Since the original BPSK can not directly applied to GenQSM MIMO systems because of the zeros in the imaginary parts, we use a $\frac{\pi}{4}$ phase-rotated BPSK as the symbol modulation for GenQSM and the rotation does not change its optimality. Using the OBSS approaches, we obtain the transmit vectors as given in Appendix B when $\delta=0$, $0.1$ and $0.3$. The comparison results in terms of minimum Euclidean distance and SER are presented in Table III and Fig.~\ref{result1c}, respectively. It is observed that from Table III that the proposed OBSS approach can also bring considerable performance improvement in maximizing the minimum Euclidean distances when applied to GenQSM systems. For $\delta=0$, $0.1$ and $0.3$, the optimized minimum Euclidean distances are almost the same.  SER comparisons in Fig.~\ref{result1c} demonstrate that when $\delta=0.3$ the systems achieves a lower SER. It is inconsistent with the minimum Euclidean distance comparisons. The reason is that the SER is determined not only by the minimum Euclidean distance but also by the other pairwise Euclidean distances (c.f., Eq. \ref{UB}). Compared with GenQSM with $\frac{\pi}{4}$-BPSK, the proposed OBSS approaches bring substantial performance improvements by about $2$ dB, $5$ dB and $9$ dB for  $\delta=0$, $0.1$ and $0.3$, respectively. This validates the performance superiority of our design in GenQSM systems.

\begin{table}[t]
\centering
\caption{Minimum Euclidean Distance Comparisons in $(3,2,2,4)$ GenQSM MIMO Systems}\label{tab3}
    \begin{tabular}{ | c | c|c|c|}
    \hline
-&$\delta=0$&$\delta=0.1$&$\delta=0.3$\\
\hline
$\frac{\pi}{4}$-BPSK&$0.7071$&$0.4954$&$0.3430$\\
\hline
OBSS&$ 1.2852$&$1.2754$&$1.2614$\\
\hline
    \end{tabular}
\end{table}

\subsection{Superiority of Proposed Designs in Closed-Loop Systems with Instantaneous CSIT}
\begin{figure}[t]
  \centering
  \includegraphics[width=0.5\textwidth]{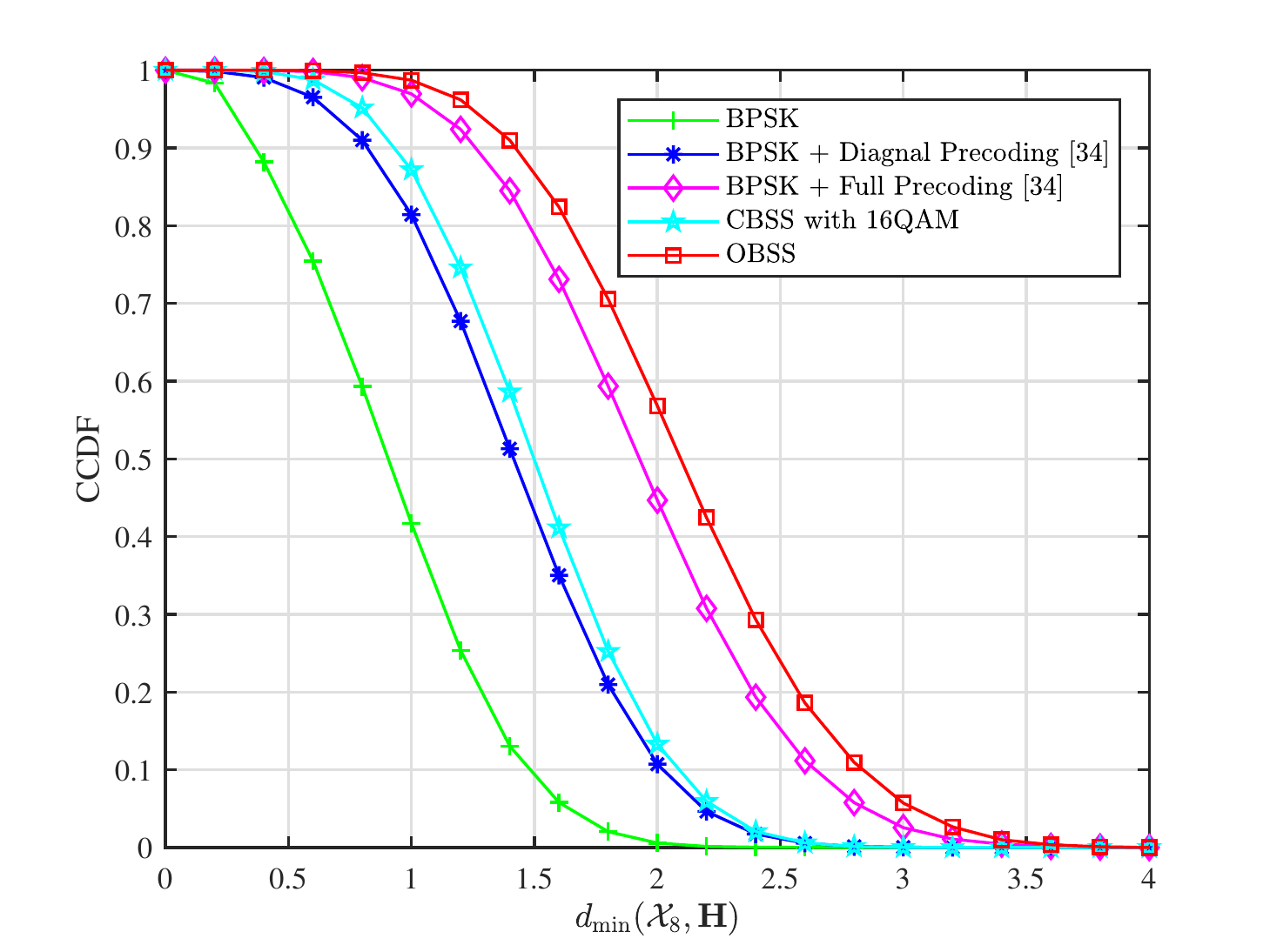}\\
 \caption{CCDF comparisons in adaptive $(3,2,2,3)$  GenSM MIMO systems.}
  \label{result2a}
\end{figure}
\begin{figure}[t]
  \centering
  \includegraphics[width=0.5\textwidth]{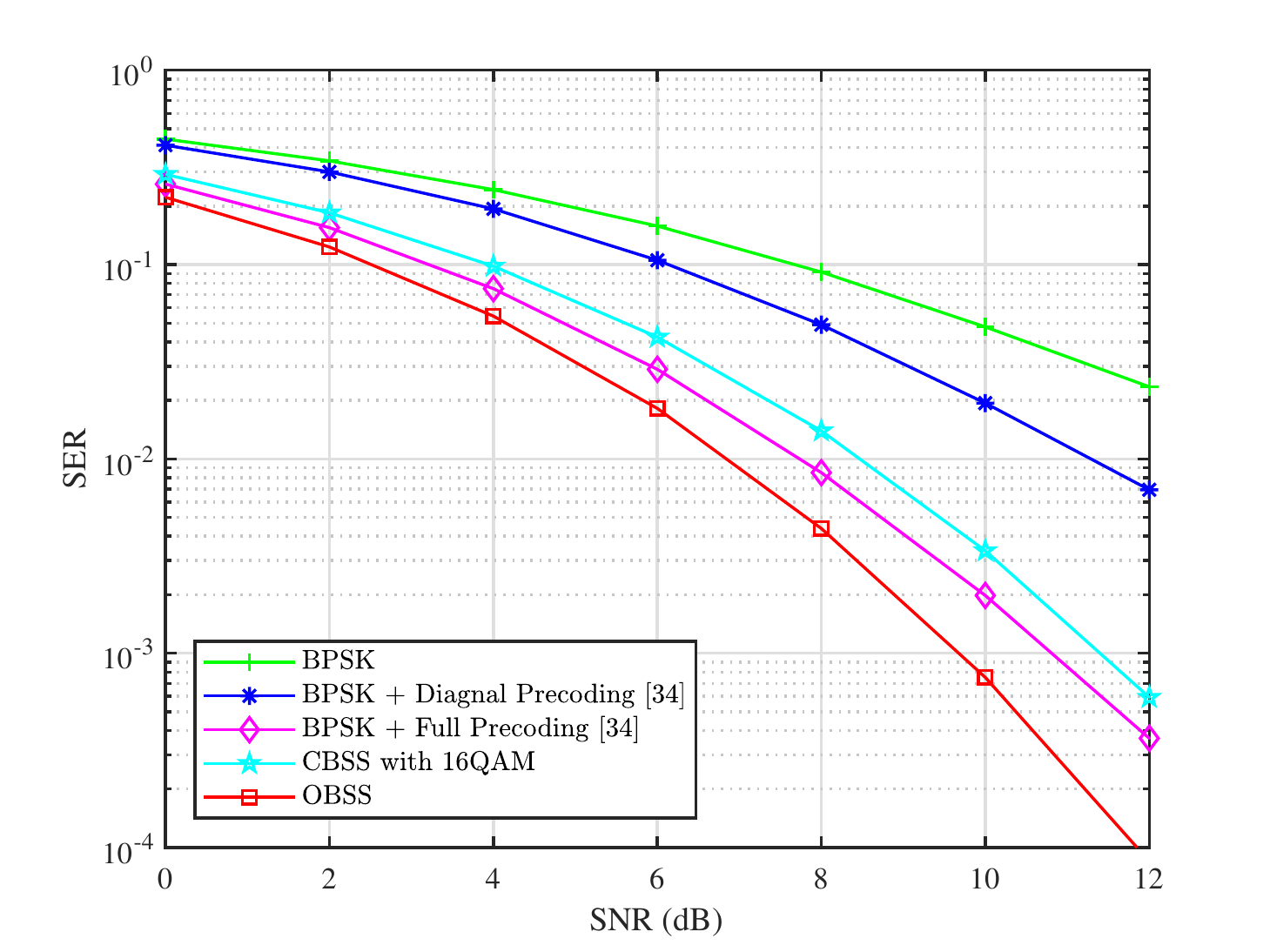}\\
 \caption{SER comparisons in adaptive   $(3,2,2,3)$  GenSM MIMO systems.}
  \label{result2b}
\end{figure}
With perfect CSIT, we investigate the proposed designs in $(3,2,2,3)$ GenSM MIMO systems, we plot the complementary cumulative distribution function (CCDF) of the minimum Euclidean distance $d_{\min}(\mathcal{X}_8,\textbf{H})$ in Fig. \ref{result2a}. The probability $\mathrm{Pr}\left\{d_{\min}(\mathcal{X}_8,\textbf{H})>1.5\right\}$ for GenSM with BPSK, GenSM with BPSK and the diagonal precoding in \cite{Cheng2018}, GenSM with BPSK with the full precoding in \cite{Cheng2018}, GenSM with CBSS and 16QAM as the codebook and GenSM with OBSS is  $0.1$, $0.43$, $0.5$, $0.8$ and $0.86$, respectively. From these results, OBSS is the best and open-loop BPSK-based shaping is the worst.  The SER comparisons among these schemes are also included as illustrated in Fig. \ref{result2b}, from which we observe that the proposed OBSS approaches outperforms BPSK with the full precoding and the one with diagonal precoding by around $1$ dB and $4$ dB at an SER of $10^{-2}$, respectively. The proposed CBSS approaches outperforms BPSK with diagonal precoding by more than $2$ dB at an SER of $10^{-2}$.  Moreover, as shown in Table I, the CBSS approach is of very low computational complexity compared to other schemes, and it would suit practical implementation in realistic MIMO systems with a limited processing power.

\subsection{Robustness to Channel Uncertainty}
\begin{figure}[t]
  \centering
  \includegraphics[width=0.5\textwidth]{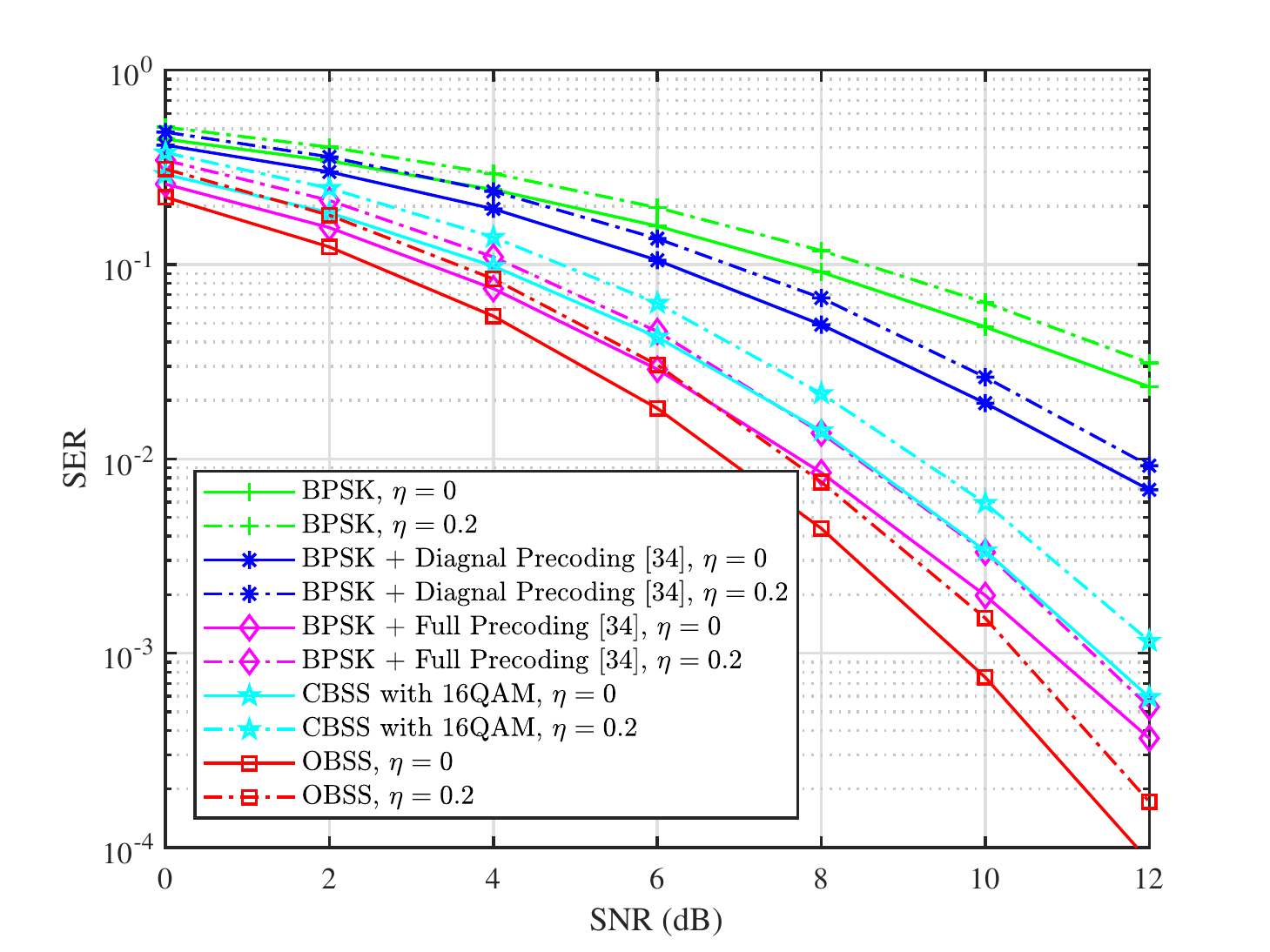}\\
 \caption{SER comparisons in adaptive  $(3,2,2,3)$  GenSM MIMO systems with/without channel estimation errors.}
  \label{result3}
\end{figure}
To show the robustness, channel estimation errors are considered and modeled as $\hat{\textbf{H}}_{\mathrm{im}}=\hat{\textbf{H}}+\hat{\textbf{H}}_{\mathrm{e}}$,
where $\hat{\textbf{H}}_{\mathrm{im}}$ denotes the estimated channel matrix, $\hat{\textbf{H}}_{\mathrm{e}}\in\mathbb{C}^{N_r\times N_t}$ represents estimation error matrix and $[\hat{\textbf{H}}_{\mathrm{e}}]_{k,l}\sim\mathcal{CN}(0,\sigma_e^2)$. In the training-based channel estimation scheme, letting $E_p$ and $N_p$ represent the average power and the number of pilot symbols, the variance $\sigma_e^2={1}/{(\rho E_pN_p)}\triangleq \eta/\rho$ if least square (LS) channel estimation is adopted \cite{Guo2016a}.
We compare the SER in presence of channel estimation errors where $\eta=0.2$ in Fig. \ref{result3}. The performance of all schemes degrades by a certain level. Despite this, the proposed OBSS approach maintains the best.
Clearly, the performance gains brought by our designs are robust to channel estimation errors.

\section{Conclusions}
This paper investigated generic signal shaping for multiple-data-steam GenSM/GenQSM MIMO systems. A unified optimization problem was formulated. We studied an OBSS approach and a CBSS approach. Results showed that the proposed approaches can be applied to open-loop systems as well as closed-loop systems to harvest a remarkable performance gain. OBSS exhibits the best performance compared to existing benchmarks. It can be used to combat the transmit correlation. Simulation results also showed that OBSS can benefit from transmit correlation to offer performance gains. CBSS shows comparable performance but only requires quite low complexity. Results  validated the superiority and the robustness of the proposed optimization strategies to channel uncertainty.

\appendices
\section{Transmit Vectors for (3,2,2,3) GenSM MIMO Systems}
When $\delta=0$, the designed transmit vectors by the proposed OBSS approach are 
\begin{equation*}
\begin{split}
&\textbf{x}_1=[-1.1177 - 0.1533j,~  0,~ -0.0007]^T,\\
&\textbf{x}_2=[0.1004 - 0.7319j,~   0,~   0.0832 - 0.5971j]^T,~\\
&\textbf{x}_3=[-0.1004 + 0.7313j,~ 0,~   0.0836 - 0.5978j]^T,~\\
&\textbf{x}_4=[0,~0.4163 + 0.4714j,~ -0.6856 + 0.2086j]^T,\\
\end{split}
\end{equation*}
\begin{equation*}
\begin{split}
&\textbf{x}_5=[0,~0.6156 - 0.3389j,~  0.5214 + 0.3777j]^T,~\\
&\textbf{x}_6=[0,~-0.5637 + 0.5430j,~   0.4065 + 0.3614j]^T,~\\
&\textbf{x}_7=[0,~-0.4699 - 0.6776j,~ -0.4120 + 0.2457j]^T,~\\
&\textbf{x}_8=[1.1178 + 0.1534j,~ -0.0001 - 0.0004j,~0]^T.
\end{split}
\end{equation*}
When $\delta=0.1$, the designed transmit vectors by the proposed OBSS approach are 
\begin{equation*}
\begin{split}
&\textbf{x}_1=[ -0.1240 - 0.8435j,~ -0.3710 - 0.4843,~   0]^T,~\\
&\textbf{x}_2=[ 0.8346 - 0.1739j,~   0.5824 + 0.1816j,~0]^T,~\\
&\textbf{x}_3=[0.1239 + 0.8434j,~   0.3710 + 0.4843j,~0]^T,~\\
&\textbf{x}_4=[  0,~   0.3067 - 0.4391j,~   0.5095 - 0.7294j]^T,~\\
&\textbf{x}_5=[ -0.3328 - 0.2325j,~0,~   0.6128 + 0.4281j]^T,~\\
&\textbf{x}_6=[ 0,~ -0.3067 + 0.4391j,~  -0.5095 + 0.7294j]^T,~\\
&\textbf{x}_7=[0.3328 + 0.2324j,~  0,~  -0.6128 - 0.4281j]^T,~\\
&\textbf{x}_8=[ -0.8346 + 0.1740j,~  -0.5824 - 0.1816j,~   0]^T.
\end{split}
\end{equation*}
When $\delta=0.3$, the designed transmit vectors by the proposed OBSS approach are 
\begin{equation*}
\begin{split}
&\textbf{x}_1=[0,~0.1419 + 0.7883j,~0.1285 + 0.7472j]^T,~\\
&\textbf{x}_2=[  0,~0.2711 - 0.9099j,~0.3027 - 1.0037j]^T,~\\
&\textbf{x}_3=[  0,~-0.5012 + 0.5787j,~-0.5744 + 0.5369j]^T,~\\
&\textbf{x}_4=[-0.7199 - 0.0137j,~-0.6850 - 0.1235j,~0]^T,~\\
&\textbf{x}_5=[-0.1523 - 0.6480j,~-0.2336 - 0.4745j,~0]^T,~\\
&\textbf{x}_6=[0,~0.6269 + 0.3302j,~0.6699 + 0.2375j]^T,~\\
&\textbf{x}_7=[0.5631 - 0.4223j,~0.4042 - 0.3051j,~0]^T,~\\
&\textbf{x}_8=[-0.0535 + 0.2969j,~0,~-0.0011 - 0.0562j]^T.
\end{split}
\end{equation*}

\section{Transmit Vectors for (3,2,2,4) GenQSM MIMO Systems}
When $\delta=0$, the designed transmit vectors by the proposed OBSS approach are 
\begin{equation*}
\begin{split}
  &\textbf{x}_1=[-0.2338 + 0.3906j,~   0.2806 - 0.8627j,~   0]^T,~\\
   &\textbf{x}_2=[0.7852 - 0.0556j,~  -0.3070 - 0.5995j,~   0]^T,~\\
  &\textbf{x}_3=[-0.4304 ,~  -0.7608 - 0.3647j,~   - 0.3581j]^T,~\\
  &\textbf{x}_4= [0,~  -0.6107 - 0.0578j,~  -0.0003 + 0.8040j]^T,~\\
  &\textbf{x}_5= [0.0068j,~   0.0017 - 0.0002j,~  -0.7925]^T,~\\
  &\textbf{x}_6= [ - 0.9776j,~  -0.1668 - 0.1926j,~  -0.0066]^T,~\\
  &\textbf{x}_7=[ 0.5328 ,~   0.8202 - 0.1502j,~   - 0.3357j]^T,~\\
  &\textbf{x}_8=[-0.0819 - 0.3075j,~   0.6306,~   0.7233j]^T,~\\
  &\textbf{x}_9=[-0.2282 + 0.5276j,~   0.5738 + 0.6377j,~   0]^T,~\\
  &\textbf{x}_{10}=[ 0.4247 ,~  -0.2670 + 0.3035j,~   - 0.8379j]^T,~\\
  &\textbf{x}_{11}=[-1.0227 + 0.1745j,~   0,~  -0.1326 + 0.3776j]^T,~\\
  &\textbf{x}_{12}=[-0.3438 - 0.2904j,~  -0.3953 + 0.8120j,~   0]^T,~\\
  &\textbf{x}_{13}=[ 0.0112,~   - 0.0035j,~   0.7889 - 0.0042j]^T,~\\
  &\textbf{x}_{14}=[-0.5899 - 0.2664j,~   0.3900,~   0.6714j]^T,~\\
  &\textbf{x}_{15}=[ 0.7613,~   0.0407 + 0.5918j,~   0.3287j]^T,~\\
    &\textbf{x}_{16}=[ 0.2101 + 0.9225j,~  -0.3728 0,~   - 0.0307j]^T
\end{split}
\end{equation*}
When $\delta=0.1$, the designed transmit vectors by the proposed OBSS approach are 
\begin{equation*}
\begin{split}
  &\textbf{x}_{1}=[-0.3765 ,~  -0.3369 - 0.5475,~    - 0.7490j]^T,~\\
  &\textbf{x}_{2}=[-0.0084 ,~  -0.2320 - 0.7504j,~    0.8066j]^T,~\\
  &\textbf{x}_{3}=[-0.7282 - 0.1672j,~  -0.6096 - 0.0433j,~    0]^T,~\\
   &\textbf{x}_{4}=[0.1880 + 0.8692j,~  -0.1510 + 0.5424j,~    0]^T,~\\
   &\textbf{x}_{5}=[0.7840 + 0.1044j,~   0.8328 ,~    - 0.1644j]^T,~\\
    &\textbf{x}_{6}=[- 0.9724j,~   0.0054 - 0.4107j,~  -0.1058 ]^T,~\\
   &\textbf{x}_{7}=[0.4920 - 0.7041j,~    0,~   0.5193 + 0.3054j]^T,~\\
  &\textbf{x}_{8}=[-0.3073 + 0.6054j,~    0,~  -0.1097 - 0.5325j]^T,~\\
   &\textbf{x}_{9}=[0.7253 + 0.0536j,~    0.2223j,~  -0.3856 ]^T,~\\
    &\textbf{x}_{10}=[0.4426j,~   0.2281 + 0.0567j,~   0.6645 ]^T,~\\
   &\textbf{x}_{11}=[0.2960 ,~   0.2766 - 0.2495j,~    - 0.6872j]^T,~\\
   &\textbf{x}_{12}=[0.1775 ,~   0.4132 + 0.4427j,~    0.8821j]^T,~\\
  &\textbf{x}_{13}=[-0.7016 ,~    0.4181j,~   0.3013 + 0.3728j]^T,~\\
  &\textbf{x}_{14}=[-0.6165 - 0.3602j,~    0,~   0.6890 - 0.2347j]^T,~\\
   &\textbf{x}_{15}=[0.4195j,~  -0.4286 ,~  -0.6306 + 0.4376j]^T,~\\
   &\textbf{x}_{16}=[0.0450 - 0.3236j,~    0,~  -0.9465 - 0.2560j]^T.
\end{split}
\end{equation*}
When $\delta=0.3$, the designed transmit vectors by the proposed OBSS approach are 
\begin{equation*}
\begin{split}
 &\textbf{x}_{1}=[0.5878 - 0.3085j,~  - 0.0947j,~-0.7883]^T,~\\
  &\textbf{x}_{2}=[-0.9532 - 0.1468j,~-0.7418 - 0.1129j,~ 0]^T,~\\
   &\textbf{x}_{3}=[0.4954 - 0.6948j,~ 0.3057 - 0.5752j,~ 0 ]^T,~\\
    &\textbf{x}_{4}=[0.6985j,~-0.4899 + 0.5915j,~-0.5052 ]^T,~\\
    &\textbf{x}_{5}=[0,~-0.4795 - 0.4245j,~-0.6073 - 0.4111j]^T,~\\
    &\textbf{x}_{6}=[0,~-0.3107 + 0.1851j,~-0.7693 + 0.0429j]^T,~\\
   &\textbf{x}_{7}=[- 0.1242j,~ 0.0515 - 0.1875j,~ 0.5202]^T,~\\
   &\textbf{x}_{8}=[0.0417 + 0.7919j,~ 0.0288 + 0.6914j,~ 0]^T,~\\
   &\textbf{x}_{9}=[0.9175,~ 0.4930 + 0.0833j,~ - 0.0945j]^T,~\\
   &\textbf{x}_{10}=[- 0.3448j,~ 0.6695 - 0.3860j,~ 0.8799 ]^T,~\\
   &\textbf{x}_{11}=[0.4810 ,~ 0.1179 + 0.2399j,~  0.4618j]^T,~\\
  &\textbf{x}_{12}=[-0.5146 ,~ 0.0645 + 0.0682j,~  0.8279j]^T,~\\
    &\textbf{x}_{13}=[0.7548j,~-0.0826 ,~-0.0552 - 0.4427j]^T,~\\
  &\textbf{x}_{14}=[-0.1837,~ 0.0788 - 0.4790j,~ - 0.6750j]^T,~\\
   &\textbf{x}_{15}=[ 0,~ 0.6779 + 0.4047j,~ 0.8375 + 0.3558j]^T,~\\
  &\textbf{x}_{16}=[-0.8045 - 0.3543j,~  0,~ 0.2353 + 0.0098j]^T.
\end{split}
\end{equation*}


\end{document}